\documentclass[%
reprint,
 amsmath,amssymb,
tightenlines
]{revtex4-2}
\usepackage{graphicx}
\usepackage{xcolor}
\usepackage{dcolumn}
\usepackage{bm}
\usepackage{setspace}
\usepackage{amsmath}
\usepackage{amssymb}
\usepackage{amsthm}
\usepackage{physics}
\usepackage{bbold}
\usepackage[T1]{fontenc}
\usepackage{mathrsfs}
\usepackage{verbatim}
\usepackage{mathtools}
\usepackage{algorithm}
\usepackage{algpseudocodex}
\usepackage[labelfont=bf]{caption}
\usepackage{hyperref}

\newcommand{\SI}[1]{}

\renewcommand{\cal}[1]{\mathcal{#1}}

\begin{document}

\title{Foundation Tensor Networks Generalize Across Lattice  Hamiltonians}
\author{Ryan T. Grimm} \email[Corresponding author: ]{ryan.grimm@berkeley.edu} \author{Michael Lindsey}  \affiliation{Department of Mathematics, University of California, Berkeley, CA 94720, USA} 

\date{\today}

\begin{abstract}

Neural quantum states and tensor networks have emerged as leading techniques for simulating quantum many-body systems. However, disordered systems remain extremely challenging in both frameworks as they require computation of the ground state for every realization of the disorder. Recently, neural quantum states have been used to construct foundation models that generate the ground state directly from the Hamiltonian. Still, they are hindered by the need to perform extensive sampling to estimate observables. To address these difficulties, we introduce the foundation tensor network (FTN), which uses a neural network to generate the tensor network representation of the ground-state wavefunction from the Hamiltonian.  To accelerate training, we develop a technique, called \texttt{GridBP}, based on iterated one-dimensional belief propagation that reduces the cost of computing the energy expectation value to just two tensor contractions. We use FTNs to calculate the critical parameters of the glass transition in the quantum Edwards-Anderson model via finite-size scaling analysis. To facilitate rapid averaging, we use a moment-generating function to evaluate the order parameter, lowering the cost to a single tensor contraction. We obtain high-fidelity estimates for the critical parameters by averaging over 500,000 disorder realizations. 

\end{abstract}

\maketitle

\section{Introduction}
Examining zero-temperature critical phenomena in disordered systems is among the most theoretically interesting problems in quantum statistical mechanics due to the complex phase behavior of these systems arising from quantum, rather than thermal, fluctuations. Spin glasses are archetypal systems in the study of disordered and glassy critical phenomena \cite{Castellani2005-vj}. In these systems, quenched disorder that persists over extremely long time scales generates a glassy phase where spins are frozen in a random but ordered state that is neither ferromagnetic nor antiferromagnetic \cite{Binder1986-fq, Edwards1975-qi}. Here, a central problem is to understand the physics of how spin glasses exit the glassy phase into other disordered phases. The Edwards-Anderson transverse-field quantum Ising (EA-TFQIM) model, examined in this paper, is a paradigmatic disordered system that undergoes a transition from a glassy to a disordered paramagnetic phase as the strength of the transverse field increases \cite{Rieger1994-ug}.

Simulating quantum spin glasses requires gathering statistics over many ground states generated by different realizations of the disorder, making it one of the most challenging problems in computational physics. The direct approach of exactly diagonalizing \cite{Lin1990-tx} the Hamiltonian is intractable for systems of even moderate size, making Monte Carlo methods the dominant technique \cite{Gonzalez-Adalid-Pemartin2024-no, Brodoloni2024-qf, Lunts2023-js, Patel2024-vm, Patel2025-sr}. One such approach \cite{Gonzalez-Adalid-Pemartin2024-no} converts the system from a $d$-dimensional quantum system to a $(d + 1)$-dimensional classical system by employing a Trotter decomposition in imaginary time \cite{Suzuki1976-li}. To understand quantum phase transitions, classical Monte Carlo methods must approach the limit of infinite inverse temperature, which imposes increasing computational costs. 

Variational Monte Carlo \cite{Sorella2005-ly, Sandvik2007-fi} (VMC), meanwhile, optimizes a trial wavefunction by drawing samples from the corresponding Born rule distribution. VMC incurs both sampling and optimization costs to converge the variational ansatz to the ground state. Applying these methods to quantum spin glasses is computationally prohibitive as they require two layers of averaging: disorder and quantum. One must first find the ground state for each realization of the disorder and then draw samples from that wavefunction to estimate observables. Neural network parametrization of the variational wavefunction within VMC defines the emerging framework of neural quantum states (NQS) \cite{Carleo2017-sc,Lange2024-hu}. Recent work shows that NQS can act as foundation models for a range of Hamiltonians \cite{Rende2025-zm}, reducing the need to optimize the trial wavefunction for each realization of the disorder. However, one must still perform extensive sampling to obtain accurate estimates of observables from an NQS. 

Tensor networks define a prominent alternative to Monte Carlo for quantum many-body simulations \cite{Orus2019-og}. They encode the wavefunction as a tensor network state (TNS), a contraction of tensor cores arranged in a network that mirrors the topology of the system they represent. The central advantage is that observables can be computed deterministically as tensor contractions. The TNS representation of a ground state can be obtained by directly optimizing the tensor cores to minimize the variational energy \cite{Liao2019-fk}, cooling the state through imaginary-time evolution~\cite{Vidal2004-kk}, or evaluating the partition function for the $(d + 1)$-dimensional classical system \cite{Xie2012-pr}. Tensor networks have proven effective for computing properties of lattice systems, such as the quantum Ising and Heisenberg models \cite{Lubasch2014-ee}, among a range of other applications \cite{Grimm2026-bu, Grimm2024-yg, Grimm2024-vq, Strathearn2018-jc, Park2024-fm, Chan2011-sb}. Provided that contraction is feasible \cite{Schuch2007-mb} for a given lattice, the primary cost of tensor networks lies in finding the ground state. Recent work has also combined tensor networks with both VMC and NQS \cite{Du2026-nt, Chen2023-su} methods. However, these approaches, although highly accurate and expressive, still employ Monte Carlo sampling to estimate observables and suffer from the usual sampling costs. 

In this paper, we introduce a methodological framework called foundation tensor networks (FTNs) that combines the strengths of NQS and TNS. FTNs learn a foundation model for the tensor network cores that generalizes across Hamiltonians. In contrast to other hybridizations of neural and tensor networks, the wavefunction returned by the FTN is fundamentally a TNS.  We use a transformer \cite{Vaswani2017-tb} to directly generate the tensor network representation of the ground state from a graph embedding of the Hamiltonian. In disordered systems, this circumvents the need to find the ground state for each sample, amortizing the optimization work. To demonstrate FTNs, we train a foundation model to produce the ground states of the EA-TFQIM as a function of the transverse field and all site-site couplings. We use this to examine the transition from the low-field glassy phase to the high-field disordered phase. Directly averaging the one-shot ground states across realizations of the disorder enables us to compute the critical parameters via finite-size scaling analysis \cite{Houdayer2004-io}. We obtain estimates that match and refine results in the literature, which have significant uncertainty, demonstrating that FTNs are a powerful new tool for exploring critical phenomena.         

\section{The Quantum Edwards-Anderson Model}
We study the critical phenomena of the EA-TFQIM on a square $L\times L$ two-dimensional lattice with $N=L^2$ sites and open boundary conditions. Its Hamiltonian takes the form \begin{equation}
    \mathcal{H}(\bm J) = \sum_{\langle a, a' \rangle} J_{a,a'} Z_a Z_{a'} + \Gamma \sum_a X_a,
    \label{eq:Ham}
\end{equation}
where $X_a$ and $Z_a$ are Pauli operators acting on spin $a$, $J_{a,a'}$ is the coupling between sites $a$ and $a'$, and $\Gamma$ is the transverse field. The first sum runs over unordered nearest-neighbor pairs. The couplings $J_{a,a'}$ are i.i.d. Gaussian random variables with unit variance and zero mean, $J_{a,a'} \sim \mathcal{N}(0, 1)$. All units in this study are non-dimensionalized in terms of a natural coupling scale that coincides with the width of the coupling distribution.

Each set of random couplings $\bm J = \{J_{a,a'}\}$ generates a different ground-state wavefunction that minimizes the Rayleigh quotient
\begin{equation}
    \ket{\Psi_{\bm J}} \in \operatorname{argmin}_{\ket{\Psi}} \frac{\bra{\Psi} \mathcal{H}(\bm J) \ket{\Psi}}{\bra{\Psi}\ket{\Psi}}. 
    \label{eq:opt}
\end{equation}
At zero temperature, the EA-TFQIM exhibits a quantum phase transition mediated by the transverse field. At $\Gamma_C$ the system transitions from a glassy phase to a paramagnetic phase. The order parameter for the transition is the replica overlap
\begin{equation}
    Q = \frac{1}{N^2} \mathbb{E}_{\bm J}\left[ \sum_{a, a'} \langle Z_{a} Z_{a'} \rangle_{\bm J}^2 \right], 
    \label{eq:overlap}
\end{equation}
where $\langle \cdot \rangle_{\bm J}$ denotes a quantum average taken with respect to the ground state with the couplings $\bm J$.
\section{Amortized Disorder Sampling}
To accelerate averaging over the disorder, we learn a mapping between the parameters of the Hamiltonian and a representation of the ground-state wavefunction. Then we can average over the disorder by directly generating wavefunctions from the disorder samples via the learned mapping. The process of learning the mapping over the entire distribution of Hamiltonians amortizes the computational effort compared to optimizing a trial state for each sample individually. We call this \textit{amortized disorder sampling}. A transformer provides an expressive ansatz to learn the mapping, and the projected entangled pair state (PEPS) enables us to efficiently represent the wavefunction and measure observables without an additional step of sampling from the Born distribution. Here, we develop a method to perform amortized disorder sampling using FTNs. 

We first expand the wavefunction in the computational basis 
\begin{equation}
    \ket{\Psi} = \sum_{\bm \sigma} \mathcal{C}_{\bm \sigma} \ket{\bm \sigma},
\end{equation}
where $\bm \sigma = (\sigma_1, \dots, \sigma_N)$ is the spin index, $\mathcal{C}$ is the coefficient tensor, and $\ket{\bm \sigma}$ is the tensor product computational basis state. Next, we encode the coefficient tensor as a projected entangled pair state (PEPS) \cite{Cirac2021-uh} of the form
\begin{equation}
    \mathcal{C}_{\bm \sigma} = \mathrm{tTr}\left[\prod_{n = 1}^N \mathcal{T}_n(\sigma_n)\right],
\end{equation}
where $\{\mathcal{T}_n\}_{n = 1}^N$ are the tensor cores, and $\mathrm{tTr}[\cdot]$ is the tensor trace that sums over virtual indices of connected cores. 

Instead of optimizing the PEPS directly, we use a transformer-based neural network to learn the mapping between Hamiltonian parameters and tensor cores. We embed information about each site, including local couplings, the transverse field, and position, into a series of $N$ tokens $\bm{\Xi} =\{ \bm{\xi}_n\}_{n = 1}^N$. After passing the tokens through the generator $\mathcal{G}_{\bm \theta}$, a series of transformer blocks, we obtain a set of latent states representing the approximate ground state of the input Hamiltonian: $\bm{\Xi}' = \mathcal{G}_{\bm \theta}(\bm{\Xi})$, where $\bm \theta$ are the parameters. The readout, $\mathcal{R}_{\bm \theta}$, a multi-layer perceptron, then converts each token into a tensor core $\mathcal{T}_n = \mathcal{R}_{\bm \theta}(\bm{\xi}'_n)$.

We construct the generator from a series of transformer blocks \cite{Vaswani2017-tb} of the form
\begin{equation}
    \mathcal{U}_{\bm \theta}(\bm{\Xi}) = \left[\left( \mathcal{P}_{\bm \theta}\circ\mathcal{N}_{\bm \theta}  + \mathcal{I} \right)\circ\left(\mathcal{A}_{\bm \theta} \circ \mathcal{N}_{\bm \theta} + \mathcal{I} \right)\right](\bm{\Xi}),
\end{equation}
where $\cal N$, $\mathcal{I}$, $\mathcal{P}$, and $\mathcal{A}$ denote layer normalization \cite{Ba2016-ji}, the identity map, a multi-layer perceptron (MLP), and multi-head self-attention (MSA), respectively. The MSA layer takes the form 
\begin{equation}
        {[\mathcal{A}_{\bm \theta}(\bm{\Xi})]_i} = \|_{h = 1}^{N_H}  \bm M_h \left[\sum_{j = 1}^N {\frac{\exp(\bm{\xi}_i^{\intercal} \bm S_h \bm{\xi}_j)}{\sum_{k = 1}^N \exp(\bm{\xi}_i^{\intercal} \bm S_h \bm{\xi}_k)}} \bm{\xi}_j\right],
\end{equation}
where $N_H$ is the number of attention heads, $\bm M_h$ and $\bm S_h$ are learnable matrices indexed for each attention head $h$, and $\|$ denotes vector concatenation. To form the generator, we compose these transformer blocks 
\begin{equation}
    \mathcal{G}_{\bm \theta}(\bm{\Xi}) = (\mathcal{U}_{\bm \theta_{N_L}} \circ \cdots \circ {\mathcal{U}_{\bm \theta_1}}) (\bm{\Xi}),
\end{equation}
where $N_L$ is the number of layers.

The combination of the transformer and tensor network forms an FTN, which we denote as $\ket{\Psi_{\bm \theta}(\bm{\Xi})}$. We train the generator and readout to minimize the average energy over a collection of Hamiltonians \cite{Rende2025-zm}
\begin{equation}
    \bm{\theta}^{\star} \in \operatorname{argmin}_{\bm \theta} \mathbb{E}_{\bm{\Xi}} \left[ \frac{\bra{\Psi_{\bm \theta}(\bm{\Xi})} \mathcal{H}(\bm J) \ket{\Psi_{\bm \theta}(\bm{\Xi})}}{\bra{\Psi_{\bm \theta}(\bm{\Xi})}\ket{\Psi_{\bm \theta}(\bm{\Xi})}} \right].
    \label{eq:multi-Ham}
\end{equation}
Once trained, the FTN enables us to compute quantities like the overlap parameter in Eq. \ref{eq:overlap} directly by evaluating the neural network and contracting the resulting PEPS. 

\begin{figure*}
    \centering
    \includegraphics[width=1.0\linewidth]{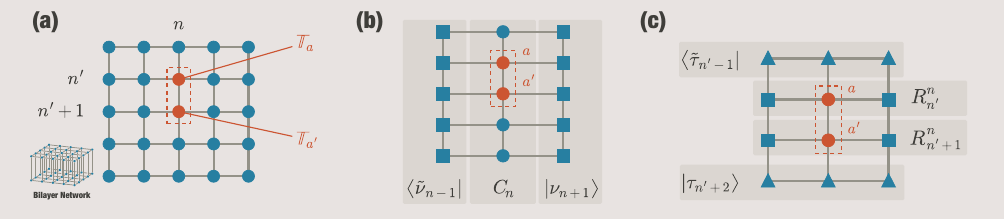}
    \caption{An illustration of the \texttt{GridBP} method for computing the two-body environment tensor. We select a pair of cores in the bilayer network \textbf{(a)} and isolate their column using a first round of message passing left and right between columns \textbf{(b)}. Then, from the MPS-MPO-MPS product, we isolate their rows via a second round of message passing up and down the network \textbf{(c)}. Finally, we differentiate the resulting $3 \times 4$ network with respect to $\mathbb{T}_a$ and $\mathbb{T}_{a'}$ to obtain the two-body environment tensor.  }
    \label{fig:gridbp}
\end{figure*}

\section{Grid Belief Propagation}
An essential ingredient in the feasibility of FTNs is fast training. At each training step, we must evaluate the energy and its gradient with respect to the parameters $\bm \theta$. Both require one or more contractions, which are significantly more expensive than performing inference on the transformer. This makes the contraction step the computationally limiting factor. Thus, we attempt to minimize the number of contraction steps. There are several ways to compute the energy expectation value with a PEPS. One is to form a projected entangled pair operator (PEPO) representation of the Hamiltonian \cite{Lubasch2014-ee}. Another is to use linearity and evaluate the Hamiltonian term by term. The former requires enlarging the bond dimension, which is undesirable due to the high cost of PEPS contraction \cite{Lubasch2014-wx}, and the latter scales quadratically with the number of sites. Inspired by belief propagation algorithms for tensor networks \cite{Alkabetz2021-ej, Evenbly2026-xh}, we derive a method called grid belief propagation (\texttt{GridBP}) that enables us to compute the energy and its gradient, via automatic differentiation, with only two contractions and no increase in the bond dimension. 

Consider a general product of states, $\bra*{\tilde{\phi}}$ and $\ket{\phi}$, and operators, $\{O_m\}_{m = 1}^M$, of the form 
\begin{equation}
   \mathcal{Z} = \matrixelement{\tilde{\phi}}{\prod_{m=1}^{M} O_m}{\phi}.
\end{equation}
By defining messages with a recursion relation of the form 
\begin{align}
     &\ket{\mu_n} = O_n\ket{\mu_{n+1}}, \ n \leq M \\
        &\bra*{\tilde{\mu}_n} = \bra{\tilde{\mu}_{n-1}}O_n, \ n \geq 1,
\end{align}
where $\ket{\mu_{M+1}} = \ket{\phi}$ and $\bra{\tilde{\mu}_0} = \bra*{\tilde{\phi}}$,
we can isolate a particular operator at position $n=1,\ldots,M$ via 
\begin{equation}
    \mathcal{Z} = \bra{\tilde{\mu}_{n - 1}}O_n\ket{\mu_{n + 1}}.
\end{equation}
We can exploit this relation to isolate pairs of sites in a PEPS contraction and obtain the two-body environment tensor
\begin{equation}
        \mathcal{E}_{a, a'} = \left( \partial_{\mathbb{T}_a} \otimes \partial_{\mathbb{T}_{a'}} \right) \mathcal{Z}_\Psi,
\end{equation}
which is sufficient to compute the energy contribution of any two-body term. Here $\mathcal{Z}_\Psi = \bra{\Psi}\ket{\Psi}$ is the bilayer network, cf. Fig. \ref{fig:gridbp} (a), and $\mathbb{T}_a = \sum_\sigma \overline{\mathcal{T}_a(\sigma)} \otimes \mathcal{T}_a(\sigma)$ is the core bra-ket pair at site $a$. We first isolate a column at $n$: 
\begin{equation}
    \mathcal{Z}_{\Psi} = \bra{\tilde{\nu}_{n-1}} C_n \ket{{\nu}_{n+1}},
\end{equation}
where $C_n$ is the matrix product operator (MPO) for column $n$ and $\bra{\tilde{\nu}_n}$ and $\ket{{\nu}_n}$ are matrix product states (MPS), cf. Fig. \ref{fig:gridbp} (b). Next, we isolate a pair of rows in the MPS-MPO-MPS product at sites $n'$ and $n' + 1$ in column $n$, Fig. \ref{fig:gridbp} (c),
\begin{equation}
    \mathcal{Z}_{\Psi} = \bra{\tilde{\tau}_{n'-1}} R^n_{n'} R^n_{n' + 1} \ket{{\tau}_{n' + 2}}.
\end{equation}
$R^n_{n'}$ and $R^n_{n' + 1}$ are the row MPOs. $ \ket{{\tau}_{n' + 2}}$ and $\bra{\tilde{\tau}_{n' - 1}}$ are the column matrix product states. Finally, we take derivatives with respect to the cores to generate the two-body environment tensor
\begin{equation}
    \mathcal{E}_{a, a'} = \bra{\tilde{\tau}_{n'-1}}\partial_{\mathbb{T}_a}  R^n_{n'} \otimes \partial_{\mathbb{T}_{a'}} R^n_{n' + 1} \ket{{\tau}_{n' + 2}},
\end{equation}
where $a = (n, n')$ and $a' = (n, n' +1)$. The one-body environment follows similarly. By iterating both recursion relations, we obtain all one-body and two-body environments along all columns. Performing the same procedure on the transposed network generates the two-body environments along all rows. Because the messages are cached in each stage, this procedure has the same cost as two contractions of the bilayer PEPS.

\section{Replica Moment-Generating Functions}
Computing the overlap parameter~\eqref{eq:overlap} requires computing correlations over all pairs of sites. Naively, this would necessitate $\mathcal{O}(N^2)$ contractions. To avoid this, we use an approach based on replica moment-generating functions (MGFs). To handle the quadratic dependence on the correlation matrix, we introduce a replica wavefunction $\ket{\hat{\Psi}} = \ket{\Psi} \otimes \ket{\Psi}$ and a Pauli-Z operator $\hat{Z}_a = Z_a \otimes Z_a$. This reduces the per-sample overlap parameter to 
\begin{equation}
    q = \frac{1}{N^2} \sum_{a, a'} \langle \hat{Z}_a \hat{Z}_{a'} \rangle_{\hat{\Psi}},
\end{equation}
where $\langle \cdot \rangle_{\hat{\Psi}}$ denotes the average taken with respect to the replica wavefunction, and $Q = \mathbb{E}_{\bm J}[q]$. Then its MGF is 
\begin{equation}
    \mathcal{M}(\bm \lambda) = \left\langle \exp(\bm \lambda^{\intercal} \hat{\bm{Z}}) \right\rangle_{\hat{\Psi}},
\end{equation}
where $\bm \lambda = \{\lambda_a\}_{a = 1}^N$ are the MGF parameters, and $\hat{\bm Z} = {\{\hat{Z}_a\}_{a=1}^N = \{Z_a \otimes Z_a\}_{a=1}^N}$ is {a vector of replica Pauli-Z products}. The overlap parameter reduces to
\begin{equation}
    q = \frac{1}{N^2} \sum_{a, a'}{\left. \partial_{\lambda_a} \partial_{\lambda_{a'}} \mathcal{M}(\bm \lambda) \right|_{\bm \lambda = \bm 0}}.
\end{equation}
We can transform this into a form that requires a single contraction using the Hessian-vector product trick:
\begin{equation}
    q = \frac{1}{N^2} \partial_{\alpha_1} \partial_{\alpha_2} \mathcal{M}([\alpha_1 + \alpha_2]\bm{1}) \big \vert_{(\alpha_1, \alpha_2) = (0, 0)},
\end{equation}
where $\bm{1}\in\mathbb{R}^N$ is the all-ones vector. This only costs one evaluation of $\mathcal{M}$ under automatic differentiation.

\begin{figure}
    \centering
    \includegraphics[width=1.0\linewidth]{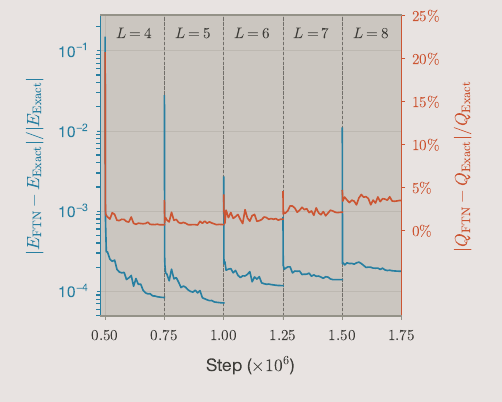}
    \caption{Convergence of energy and overlap parameter error during training. ``Steps'' denotes the total number of steps accumulated across all training phases. Dotted vertical lines denote transfer learning to a new system size. }
    \label{fig:train}
\end{figure}

\section{Training and Validation}
\begin{figure*}
    \centering
    \includegraphics[width=1.0\linewidth]{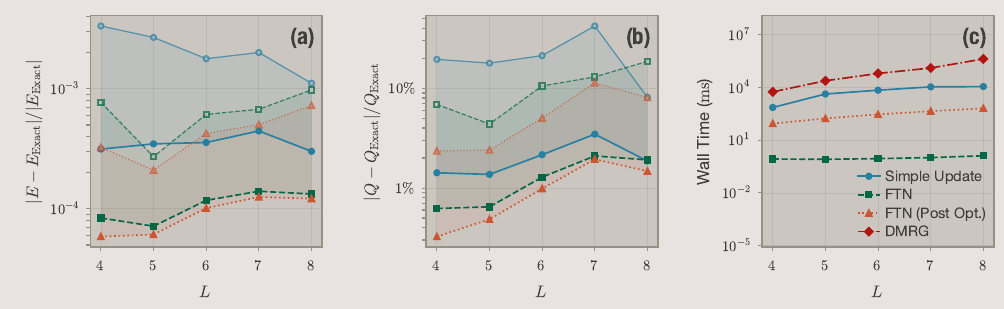}
    \caption{Comparison of FTNs (with and without post-inference optimization), DMRG, and simple update (SU). Mean and maximum errors are denoted by closed and open markers, respectively.  \textbf{(a)} Energy error vs. system size. \textbf{(b)} Overlap parameter~\eqref{eq:overlap} error vs. system size. \textbf{(c)} Wall time to compute the approximate ground state.  }
    \label{fig:bench}
\end{figure*}
To train the FTN, we perform transfer learning across system sizes from $L = 2$ to $L=10$. At each stage, the FTN is exposed only to systems of the new size. We use a single architecture for all sizes, with four transformer blocks of embedding width $64$, eight attention heads, and an MLP with a single hidden layer of width $256$ and GELU activation in each transformer block, and a readout MLP with two hidden layers of width $128$ and $\tanh$ activations. At each size, we optimize the multi-Hamiltonian objective in Eq. \ref{eq:multi-Ham} using 250,000 steps of \texttt{Adam} \cite{Kingma2014-ob}  with a cosine annealing \cite{Loshchilov2016-sc} schedule from $10^{-3}$ to $10^{-6}$. 

For each step, we draw a new set of fields from a uniform distribution $\Gamma \sim \mathcal{U}(1.5, 2.5)$ and disorder samples from the normal distribution mentioned above with a batch size of four. We compute the gradient with respect to the parameters via automatic differentiation of the energy computed by \texttt{GridBP}. In all cases, we use a bond dimension of $\chi = 2$ and a bond dimension cutoff of $\chi_{\mathrm{max}} = 16$ during contraction. To lower computational costs and avoid instabilities in the SVD during automatic differentiation \cite{Liao2019-fk}, we use randomized QR-based rounding \cite{Al_Daas2023-md}. We find that each successive size benefits significantly from the last and that only a few steps of additional training are needed to bring the error into an acceptable range, cf. Fig. \ref{fig:train}. This demonstrates the ability of the FTN to generalize across system sizes.  We compare FTN results with exact diagonalization results at $L = 4$ and with DMRG results for $5 \leq L \leq 8$. We performed the entire training run on a single Nvidia A100 GPU in approximately $52$ hours.  

We validate the FTN model by comparing it to brute-force DMRG \cite{White1992-mg}, as well as imaginary-time evolution using simple update (SU) \cite{Lubasch2014-ee}. We run SU with a decreasing time-step schedule from $0.1$ to $10^{-7}$ and values of $\chi$ and $\chi_{\mathrm{max}}$ matching those of the FTN. At each time-step size, convergence is determined by an absolute change in energy per site below $10^{-8}$ for three consecutive checks, performed every 100 evolution steps. Similarly, we run DMRG sweeps at a maximum bond dimension of $\chi_{\mathrm{max}} = 1024$, with convergence determined by an absolute change in total energy below $10^{-11}$ between successive sweeps. For the FTN, we compare the ground states generated directly from inference and after ten additional fine-tuning steps of energy optimization at a learning rate of $10^{-3}$. We measure the time required for each algorithm to generate its approximate ground state. We find that the FTN without post-inference optimization is as accurate as or more accurate than SU in almost all cases, cf. Fig.~\ref{fig:bench}(a),(b), and is also faster, cf. Fig.~\ref{fig:bench}(c). Moreover, we observe that post-inference optimization of the FTN output does not significantly improve results, suggesting that the PEPS output by the FTN for each disorder realization is already near-optimal for its bond dimension.

\begin{figure}
    \centering
    \includegraphics[width=1.0\linewidth]{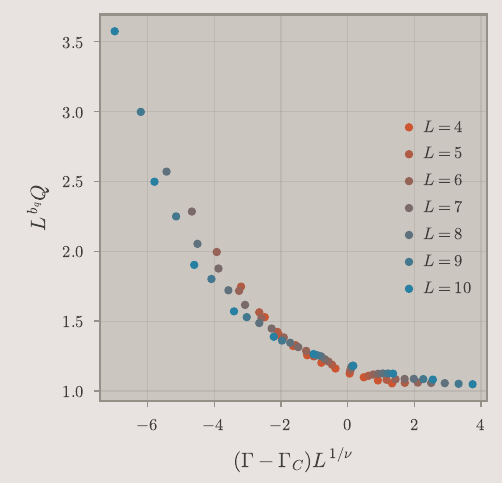}
    \caption{Collapse of the rescaled overlap parameter to a single curve under finite-size scaling. }
    \label{fig:placeholder}
\end{figure}
\begin{table}
    \caption{Critical parameters of the glass transition obtained from finite-size scaling analysis.}
    \label{tab:fss-fit-parameters}
    \begin{ruledtabular}
        \begin{tabular}{ccc}
            Parameter & Estimate & 95\% CI \\
            \hline
            $\Gamma_C$ & $2.121$ & $[2.115,\,2.126]$ \\
            $1/\nu$ & $1.128$ & $[1.104,\,1.151]$ \\
            $b_q$ & $1.918$ & $[1.914,\,1.921]$ \\
        \end{tabular}
    \end{ruledtabular}
\end{table}

\section{The Glass Transition}
To study the critical phenomena of the EA-TFQIM, we use our FTN model to compute over 500,000 ground states for different realizations of the disorder. This was possible using eight A6000 GPUs in roughly $24$ hours since our combined methodology is able to generate a PEPS, fine-tune it, and measure its overlap parameter (Eq. \ref{eq:overlap}) in under three seconds per sample. Indeed, we compute each overlap parameter using the replica MGF method, applied to the FTN output after post-inference optimization.

The transition is characterized by three critical parameters: $\Gamma_C$, the critical field at which the transition occurs; $\nu$, the scaling exponent of the correlation length; and $b_q$, the scaling exponent of the order parameter. We use finite-size scaling (FSS) analysis to determine the critical parameters, cf. Fig. \ref{fig:placeholder}. FSS posits that the rescaled curves $L^{b_q}Q(\Gamma,L)$ should collapse to a single curve when plotted against $S(\Gamma,L)$. A widely used FSS ansatz for the curve is
\begin{equation}
    Q(\Gamma, L) = L^{-b_q}\left[c_0 + c_1 S(\Gamma, L) + c_2 S^2(\Gamma, L) \right],
\end{equation}
under the transformation 
\begin{equation}
    S(\Gamma, L) = (\Gamma - \Gamma_C)L^{1/\nu},
\end{equation}
where $c_0$, $c_1$, and $c_2$ are coefficients. We fit the curves using nonlinear least squares from $128$ random starts  and estimate the $95\%$ confidence intervals (CIs)  by drawing $1024$ bootstrapped samples with replacement.

Our results, reported in Table~\ref{tab:fss-fit-parameters}, are in close agreement with other studies of the EA-TFQIM \cite{Brodoloni2024-qf, Gonzalez-Adalid-Pemartin2024-no, King2023-hn}. The large number (7168) of samples per point enables us to obtain narrow CIs quantifying uncertainty due to disorder under our model assumptions. 
\section{Acknowledgments }
This work was supported by the U.S. Department of Energy, Office of Science, Accelerated Research in Quantum Computing Centers, Quantum Utility through Advanced Computational Quantum Algorithms, grant no. DE-SC0025572.

\newpage
\clearpage

\bibliography{references}

\end{document}